\documentclass[12pt,preprint]{aastex}

\shorttitle{SDSS QUASAR LENS SEARCH. IV.}
\shortauthors{INADA ET AL.}

\begin{document}

\title{The Sloan Digital Sky Survey Quasar Lens Search. IV. \\
Statistical Lens Sample from the Fifth Data Release}

\author{
Naohisa Inada,\altaffilmark{1,2} 
Masamune Oguri,\altaffilmark{3,4} 
Min-Su Shin,\altaffilmark{5,6} 
Issha Kayo,\altaffilmark{7,8} 
Michael A. Strauss,\altaffilmark{6} 
Joseph F. Hennawi,\altaffilmark{9,10}
Tomoki Morokuma,\altaffilmark{8,11} 
Robert H. Becker,\altaffilmark{12,13} 
Richard L. White,\altaffilmark{14}
Christopher S. Kochanek,\altaffilmark{15} 
Michael D. Gregg,\altaffilmark{12,13} 
Kuenley Chiu,\altaffilmark{16} 
David E. Johnston,\altaffilmark{17} 
Alejandro Clocchiatti,\altaffilmark{18}
Gordon T. Richards,\altaffilmark{19}
Donald P. Schneider,\altaffilmark{20} 
Joshua A. Frieman,\altaffilmark{21,22,23}
Masataka Fukugita,\altaffilmark{7,24,25}
J. Richard Gott, III,\altaffilmark{6}
Patrick B. Hall,\altaffilmark{26} 
Donald G. York,\altaffilmark{23,27}
Francisco J. Castander,\altaffilmark{28} 
and
Neta A. Bahcall\altaffilmark{6}
}

\altaffiltext{1}{Cosmic Radiation Laboratory, RIKEN, 2-1 Hirosawa, Wako,
                 Saitama 351-0198, Japan.} 
\altaffiltext{2}{Research Center for the Early Universe, School of Science, 
                 University of Tokyo, Bunkyo-ku, Tokyo 113-0033, Japan. }                  
\altaffiltext{3}{Division of Theoretical Astronomy, National
                 Astronomical Observatory, 2-21-1, Osawa,
                 Mitaka, Tokyo 181-8588, Japan.} 
\altaffiltext{4}{Kavli Institute for Particle Astrophysics and
                 Cosmology, Stanford University, 2575 Sand Hill Road,
                 Menlo Park, CA 94025, USA.}                
\altaffiltext{5}{Department of Astronomy, University of Michigan, 
                 500 Church Street, Ann Arbor, MI 48109-1042 USA.}
\altaffiltext{6}{Princeton University Observatory, Peyton Hall,
                 Princeton, NJ 08544, USA.}                                   
\altaffiltext{7}{Institute of the Physics and Mathematics of the Universe, 
                 University of Tokyo, Kashiwa, 277-8582, Japan.}
\altaffiltext{8}{Research Fellow of the Japan Society for the Promotion of Science.}
\altaffiltext{9}{Department of Astronomy, University of California 
                 Berkeley, 601 Campbell Hall, Berkeley, CA
                 94720-3411, USA.}
\altaffiltext{10}{Max Planck Institut f\"{u}r Astronomie, 
                 K\"{o}nigstuhl 17, 69117 Heidelberg, Germany.}                    
\altaffiltext{11}{Optical and Infrared Astronomy Division, National
                 Astronomical Observatory, 2-21-1 Osawa, Mitaka,  
                 Tokyo 181-8588, Japan.}
\altaffiltext{12}{IGPP-LLNL, L-413, 7000 East Avenue, Livermore, CA 94550, USA.}
\altaffiltext{13}{Department of Physics, University of California 
                 Davis, 1 Shields Avenue, Davis, CA 95616, USA.}  
\altaffiltext{14}{Space Telescope Science Institute, 3700 San Martin
                 Drive, Baltimore, MD 21218, USA.} 
\altaffiltext{15}{Department of Astronomy, The Ohio State University, 
                  Columbus, OH 43210, USA.}                 
\altaffiltext{16}{School of Physics, University of Exeter, Stocker Road, 
                  Exeter EX4 4QL, UK.}
\altaffiltext{17}{Jet Propulsion Laboratory, 4800 Oak Grove Drive,  
                  Pasadena CA, 91109, USA.}
\altaffiltext{18}{Departamento de Astronom\'{i}a y Astrof\'{i}sica,
                  Pontificia Universidad Cat\'{o}lica de Chile, Casilla
                  306, Santiago 22, Chile.}
\altaffiltext{19}{Department of Physics, Drexel University, 3141
                  Chestnut Street,  Philadelphia, PA 19104, USA.}
\altaffiltext{20}{Department of Astronomy and Astrophysics, The
                  Pennsylvania State University, 525 Davey Laboratory, 
                  University Park, PA 16802, USA.}   
\altaffiltext{21}{Kavli Institute for Cosmological Physics, University 
                  of Chicago, Chicago, IL 60637, USA.}
\altaffiltext{22}{Center for Particle Astrophysics, Fermi National 
                  Accelerator Laboratory, P.O. Box 500, Batavia, IL 60510, USA.}
\altaffiltext{23}{Department of Astronomy and Astrophysics, The University 
                  of Chicago, 5640 South Ellis Avenue, Chicago, IL 60637, USA.}
\altaffiltext{24}{Institute for Cosmic Ray Research, University of Tokyo, 
                  Kashiwa, 277-8582, Japan.} 
\altaffiltext{25}{Institute for Advanced Study, Princeton, NJ08540, USA.}
\altaffiltext{26}{Department of Physics and Astronomy, York University,
                  4700 Keele Street, Toronto, Ontario, M3J 1P3, Canada.}
\altaffiltext{27}{Enrico Fermi Institute, The University of Chicago,
                  5640 South Ellis Avenue, Chicago, IL 60637, USA.}                  
\altaffiltext{28}{Institut de Ci\`encies de l'Espai (IEEC/CSIC), Campus UAB, 
                  08193 Bellaterra, Barcelona, Spain.}
  
\begin{abstract}
We present the second report of our systematic search for strongly
lensed quasars from the data of the Sloan Digital Sky Survey (SDSS).
From extensive follow-up observations of 136 candidate objects, we
find 36 lenses in the full sample of 77,429 spectroscopically
confirmed quasars in the SDSS Data Release 5. We then define a
complete sample of 19 lenses, including 11 from our previous search
in the SDSS Data Release 3, from the sample of 36,287 quasars
with $i<19.1$ in the redshift range $0.6<z<2.2$, where we require the
lenses to have image separations of $1''<\theta<20''$ and $i$-band
magnitude differences between the two images smaller than $1.25$~mag.
Among the 19 lensed quasars, 3 have quadruple-image configurations,
while the remaining 16 show double images. This lens sample
constrains the cosmological constant to be  
$\Omega_\Lambda=0.84^{+0.06}_{-0.08}({\rm stat.})^{+0.09}_{-0.07}({\rm syst.})$ 
assuming a flat universe, which is in good agreement with 
other cosmological observations. We also report the discoveries of 7 
binary quasars with separations ranging from 1\farcs1 to 16\farcs6, 
which are identified in the course of our lens survey. This study 
concludes the construction of our statistical lens sample in the full 
SDSS-I data set.   
\end{abstract}

\keywords{gravitational lensing: strong --- quasars: general --- 
          cosmology: observations}
                   
\section{INTRODUCTION}\label{sec:intro}

Gravitationally lensed quasars are useful tools for a variety of
astrophysical and cosmological studies
\citep[e.g.,][]{turner84,blandford87,schneider92,kochanek06}. In 
particular, statistical analyses of lensed quasars serve as a useful
probe for the cosmological constant \citep{fukugita90,turner90} and
the Hubble constant \citep{oguri07a}. The Hubble Space Telescope ({\em
  HST}) Snapshot survey \citep{maoz93} and the Cosmic-Lens All Sky
Survey \citep[CLASS;][]{myers03,browne03} have provided examples of
complete lens samples that have allowed cosmological studies. The {\em
  HST} Snapshot survey includes five lenses selected from 502 bright, 
relatively high-redshift quasars, and was used to derive a limit on
the cosmological constant \citep{maoz93}. It was also used to study
galaxy evolution by combining with other lens surveys
\citep{chae10}. CLASS contains 13 lenses from 8958 radio sources; the
redshift distribution of this sample is not well determined
\citep[e.g.,][]{munoz03}. This sample has also been used to constrain
cosmological models as well as the structure and evolution of lens
galaxies \citep[e.g.,][]{rusin01,mitchell05,chae06}. While large
samples of galaxy-galaxy lenses are being assembled by various groups 
\citep[e.g.,][]{bolton06,bolton08,cabanac07,faure09,marshall09,kubo09,feron09}
and are used to study the structure of the lens galaxies, they are
not well suited as a cosmological probe because 
galaxy-galaxy lenses are often selected from samples of lensing 
objects, not source objects, much complicating the statistics. 

In order to construct a large statistical sample of lensed quasars
that can be used as a cosmological probe, we have conducted the Sloan
Digital Sky Survey Quasar Lens Search \citep[SQLS;][hereafter Paper
I]{oguri06} based on a sample of spectroscopically confirmed quasars
\citep{schneider07} derived from the Sloan Digital Sky Survey
\citep[SDSS;][]{york00}.  To allow cosmological tests, the lens sample
must be complete with well-defined criteria. We refer to a sample that
allows statistical tests as a ``statistical sample''.  Our 
lens sample is, in fact, designed to be complete under prescribed
conditions and therefore suitable for statistical studies and cosmological
tests, given the accurately defined selection function (see Paper I) and 
the homogeneity of the SDSS data.  In \citet{inada08}, hereafter Paper II,
we presented a complete sample of 11 lensed quasars suitable for
statistical analyses, selected from the 22,683 quasars satisfying
$0.6<z<2.2$ and $i<19.1$\footnote{Here $i$ is the Point Spread
  Function (PSF) magnitude corrected for Galactic extinction from the
  maps of \citet{schlegel98}.}  out of the total of 46,420 quasars
\citep{schneider05} in the SDSS Data Release 3 
\citep[DR3;][]{abazajian05}. This sample derived from the DR3 gives
cosmological constraints \citep[][hereafter Paper III]{oguri08a}
that agree with the current cosmological model
\citep[e.g.,][]{komatsu09,tegmark06}.

In this paper, we extend our source population to the SDSS Data
Release 5 \citep[DR5;][]{adelman07}, concluding the SDSS-I (the 
first phase of the SDSS project through 2005 June). The selection
process is the same as that used to create the DR3 statistical sample
of lenses, as described in Paper II. Note that all lensed quasars
described in this paper have already been reported in their discovery
papers. The main focus of this paper is to define the DR5 statistical  
sample, and to report quasars that turned out not to be gravitational 
lenses. We briefly describe the source quasar sample from which we 
constructed the DR5 statistical lens sample in Section~\ref{sec:source}, 
and the selection of lens candidates in Section~\ref{sec:cand}. We present 
observational results for the lensing candidates in Section~\ref{sec:obs}, 
and the resulting DR5 lensed quasar sample and updates of cosmological 
constraints in Section~\ref{sec:dr5_lens}. Section~\ref{sec:summary} 
gives a summary of our results. 

\section{SOURCE QUASARS}\label{sec:source}

The SDSS is a combination of photometric and spectroscopic surveys of
a quarter of the sky, primarily in a region centered on the North
Galactic Cap. The surveys were carried out using a dedicated
wide-field 2.5-m telescope \citep{gunn06} at the Apache Point
Observatory in New Mexico, USA. The details of the photometric survey
using five broad-band optical filters ($ugriz$), including the
astrometric accuracy and photometric zero point accuracy, are described
in \citet{fukugita96}, \citet{gunn98}, \citet{hogg01}, \citet{smith02}, 
\citet{pier03}, \citet{ivezic04}, \citet{tucker06}, and
\citet{padmanabhan08}. The data of the photometric survey are
processed by the photometric pipeline \citep{lupton01}. The target
selection pipeline \citep{richards02} finds quasar candidates for the
spectroscopic survey. The candidates are tiled on each plate according
to the algorithm of \citet{blanton03}. Spectroscopic observations with
a resolution of $\hbox{R}\sim1800$ are carried out using a pair of
multi-fiber double spectrographs covering 3800{\,\AA} to 9200{\,\AA}. 
The SDSS data have been published in series of Data Releases
\citep{stoughton02,abazajian03,abazajian04,abazajian05,abazajian09,
adelman06,adelman07}.

We construct a source quasar sample following the procedure in Paper
II. We start with the 77,429 quasars in the DR5 spectroscopic survey
selected over an area of 5740~${\rm deg^2}$ \citep{schneider07}, and
restrict the redshift range to $0.6<z<2.2$ and magnitude to $i<19.1$,
over which the quasar target selection is almost complete (see Paper I) 
and does not significantly bias our lens candidate search. 
The lower redshift limit is imposed to
eliminate quasars associated with resolved host galaxies, which
otherwise would dominate our candidate list, based in part on the
extent of the optical image. We further exclude quasars in poor seeing
fields, {\tt PSF\_WIDTH}$>1\farcs8$, where identification of close
pairs becomes difficult. These selection criteria give a sample
of 36,287 quasars\footnote{See the SQLS webpage
  (http://www-utap.phys.s.u-tokyo.ac.jp/\~{}sdss/sqls/) for the list
  of the 36,287 quasars.}. Removing DR3 quasars that have already been
studied in Paper II, we are left with 13,636 quasars\footnote{The
  number of DR3 quasars in this paper (22,651) differs slightly from
  that given in Paper II (22,683). 32 quasars in the DR3 are given
  different parameters (magnitudes) in the DR5 catalog and no longer meet
  the criteria for our source sample. This does not affect the number
  of lenses or lens candidates in the DR3.} 
to be studied in this work.  

\section{LENS CANDIDATES}\label{sec:cand}

Our SQLS selection procedure to identify lensed quasar candidates uses
two different algorithms; ``morphological selection'' to find
small separation lensed quasars ($\theta\lesssim 2\farcs5$) which are
not deblended into multiple components by the SDSS photometric
pipeline, and ``color selection'' to find large separation lenses
($\theta\gtrsim 2\farcs5$) whose lensed components are deblended and
thus registered as separate objects in the SDSS image catalog. 

In brief, morphological selection finds quasars that appear as 
single objects but are not well fit with the PSF. This is represented
by the parameter {\tt star\_L} for each color band which gives the
logarithmic likelihood that the object is fit by the PSF. The values 
used in the {\tt star\_L} cut (see Paper II) were tested against 
many simulated lensed quasars, as described in Paper I. We then 
fit each system to a model with two PSFs using GALFIT \citep{peng02}, 
to exclude objects that are single quasars, or obvious quasar-star or
-galaxy associations. If these objects are fit with two PSFs, the
GALFIT for single filter images results in either very small
separations of the two centroids, or very large magnitude differences
in the two components: with data in several bands, the two decomposed
objects may become mutually inconsistent for their positions and/or
colors. Therefore, we can select lens candidates by choosing objects
whose $u$- and $i$-band image fits are both ``normal'' and are in good
positional agreement (see, more quantitatively, Equation 16 of Paper
I). In practice, it is sufficient to use the $u$-band and one of the
$gri$-band images for objects at $z<2.2$. The candidates which
survive the GALFIT selection are visually inspected to exclude objects
that are clearly superpositions of a quasar and a galaxy, particularly
using their bluer band images where signals from quasar components 
($z<2.2$) are much more dominant than those from galaxies. Of the
13,636 quasars, 50 morphologically-selected candidates survive this
selection. This selection is designed to pick up all lenses that
satisfy the relative brightness criterion ($|{\Delta}i|<1.25$; see  
below) with separations between $1''$ and $2''$.
The 50 candidates we obtained are listed in Table~\ref{tab:can_mor}.

The color selection algorithm applies to the case where the lensed
images are deblended by the photometric pipeline. We search for
objects around each quasar having colors similar to those of
the quasar.  To obtain a complete sample with image separations
$1''<\theta<20''$ and $i$-band magnitude differences between
two images $|{\Delta}i|<1.25$, we search for objects with 
$\theta<20\farcs1$ and $|{\Delta} i|<1.3$ allowing for some tolerance. 
Next we exclude some candidates whose optical and radio flux ratios 
are inconsistent between the two components \citep{kochanek99}. 
We used the data from the Faint Images of the Radio Sky at Twenty 
centimeters survey \citep[FIRST;][]{becker95} for candidates with 
image separations larger than $6''$, which is the resolution of FIRST. 
We also reject
some obvious quasar-galaxy pairs by visual inspection. We exclude
low-redshift and large-separation pairs with no detectable lensing
objects in the SDSS image, because a standard galaxy model predicts
that at least one of the member galaxies of the putative lens
group/cluster should be detectable for such lens events (see Paper II
for quantitative details).  We consider all lens candidates with
separations larger than the minimum deblending separation (${\sim}1\farcs5$)
of the SDSS photometric pipeline, amounting to 88 candidates from the
13,636 quasars. Two objects out of the 88 candidates are selected by
both the morphological and color selection algorithms, and therefore
86 candidates are listed in Table~\ref{tab:can_col}. The numbers
of objects selected by the two algorithms are summarized in
Table~\ref{tab:cand}.   

\section{OBSERVATIONS}\label{sec:obs}

\subsection{Summary of Follow-up Observations}\label{sec:followup}

The 136 lensed quasar candidates taken from the 13,636 quasars
constitute our targets for follow-up observations, but we remove
two candidates that have already been examined. One is 
the morphologically selected candidate SDSS~J111816.95+074558.1 which
is the well-known lensed quasar PG1115+080 \citep{weymann80}. The
other is the color selected candidate SDSS~J165502.02+260516.5, which
is a pair of quasars with slightly different redshifts from
SDSS spectroscopy. The remaining 134 candidates were observed as
described in what follows using various telescopic facilities. The
observations include optical and/or near-infrared imaging and optical
spectroscopy, and are tabulated in Table~\ref{tab:can_mor} and
Table~\ref{tab:can_col}.   

\begin{figure}
\epsscale{.7}
\plotone{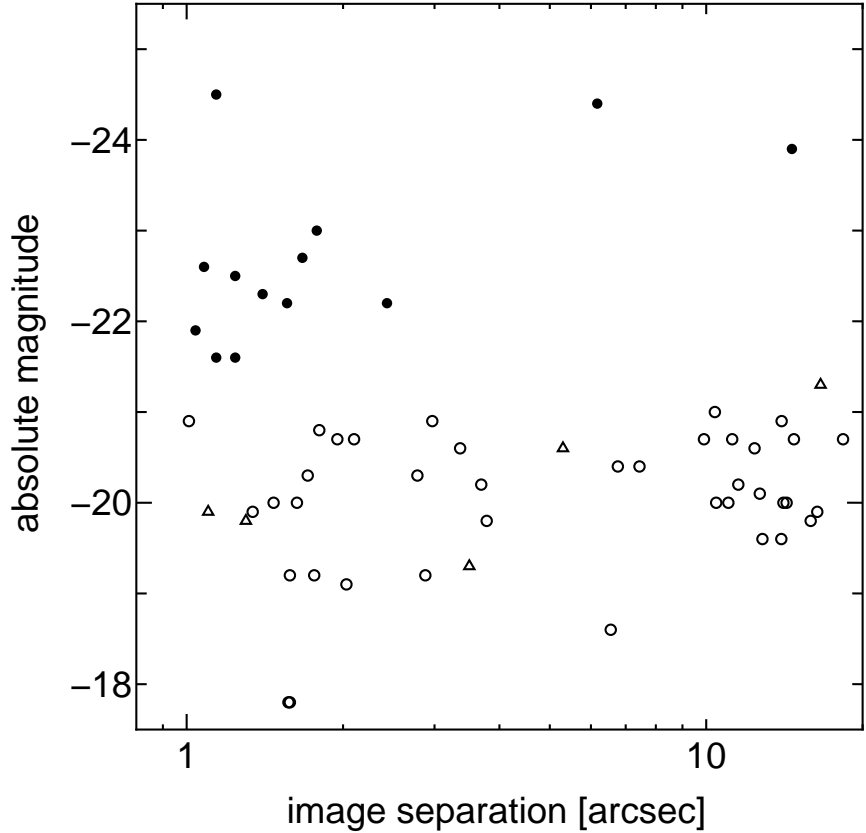}
\caption{The absolute magnitudes ($M_I$ or $M_i$) of lens galaxies 
in our DR5 statistical sample ({\it filled circles}; 13 systems with
known lens redshifts, listed in Table~\ref{tab:lens_dr5stat}) are
compared with detection limits of putative lens galaxies ({\it open
  circles} and  {\it open triangles}) for our lens candidates with
comments of ``no lens object'' and/or ``binary QSO''. We concluded
these lens candidates are not lensed quasars given the absence of lens
galaxies to our detection limits. The absolute magnitudes for the
putative lens galaxies are computed assuming ${z_l}=0.5{z_s}$ ({\it
  open circles}). For the five binary systems (Section~\ref{sec:note}), we
assumed an extreme case of $z_l=z_s$ ({\it open triangles}). The
combinations of evolution- and $K$-corrections derived from the
spectral model of \citet{poggianti97} are included. \label{fig:l-angle}} 
\end{figure}

For the 49 morphologically-selected candidates after excluding 
SDSS~J111816.95+074558.1, we first carry out optical ($i$ or
$I$) and/or infrared ($H$, $K$, or $K'$) imaging under good seeing
conditions ($\sim 0\farcs5-1\farcs0$) to confirm that the candidates
indeed have two or more stellar components and also a lens galaxy
between the stellar components.  We set the exposure time such that we
can detect faint extended objects down to $I \sim 23.0$, $H \sim
18.5$, and $K \sim 20.0$ at S/N$\simeq10$, corresponding to the
brightness at which one can detect lens galaxies if they are 
located at a redshift half that of the source quasar. 
Figure~\ref{fig:l-angle} shows the relation between the image 
separation and the rest-frame $I$ (or $i$) band luminosity of the
lens galaxies. The 13 filled circles are the luminosities of the lens
galaxies that are included in our statistical sample and have 
measured redshifts (see Table~\ref{tab:lens_dr5stat}). Open circles
are upper limit on the luminosities of hypothetical lens galaxies of the
candidates with ``no lens object'' (but not ``binary
QSO'') in Table~\ref{tab:can_mor} or Table~\ref{tab:can_col},  
assuming that the lens galaxy is located at half the source redshift. 
The assumption is reasonable given that the maximum value of the ratio 
of lens to source redshifts $z_l/z_s$ in our statistical sample is 0.60, 
and that 90\% of the lens galaxies are located at $z_l/z_s<0.5$. 
Open triangles refer to binary quasars which we discuss at length in 
Section~\ref{sec:note}, where a more extreme lens redshift $z_l=z_s$ is 
assumed. The figure indicates that the luminosities of candidate lens 
galaxies with no detection are significantly fainter than what would be 
expected for a given separation angle expected from our confirmed
lensing events. This is particularly true if we consider the fact
that a positive correlation is expected between luminosity and the
separation angle, as simple lens models predict. The empirical
correlation between the mass and luminosity of early-type galaxies
also predicts the minimum apparent luminosities of lens galaxies
\citep[e.g.,][]{rusin03}, and our detection limit corresponds  
to luminosities typically much fainter than this predicted minimum.  
For this reason we exclude as lensing candidates those cases which
do not exhibit any signatures of lens galaxy in our follow-up images. 
However, we obtained follow-up spectroscopy for four candidates which 
were judged to be particularly good candidates based on the color and 
morphology of their SDSS images, even though our follow-up images for 
these candidates show no sign of lensing objects.

Some candidates were rejected because they turn out to be single
quasars or quasar-galaxy associations. For candidates that are not
excluded at this step, we acquire spectra of the stellar components. 
Of the seven targets with possible lensing objects, five of them are
lensed quasars (six in the list of morphologically selected
systems when PG1115+080 is included). The other two objects were found
to be binary quasars (see Section~\ref{sec:note}). The four additional 
spectroscopic targets described above were found not to be lensed 
quasars. 

For the 85 color-selected candidates (after excluding SDSS~J165502.02+260516.5),
we carry out either optical/near-infrared imaging or optical
spectroscopy of the stellar components. For imaging we first looked
for archival data from the Subaru telescope \citep[SMOKA;][]{baba02},
and found that SDSS~J134150.74+283207.9 could be rejected due to the
absence of any possible lensing objects in the deep Subaru image. 
Follow-up imaging of 50 candidates yields five cases that indicate
signatures of possible lens galaxies. Subsequent spectroscopy of these
five objects shows that four of them are lensed quasars, and the
remaining one, SDSS~J160614.69+230518.0, is a quasar-star pair. We
carried out spectroscopy of the remaining 34 candidates without
imaging, which yielded five pairs of quasars with the same redshifts. 
Two of the five were found not to be lensed quasars because they have
different spectral energy distributions not consistent with being lensed 
pairs. The other three pairs were
also rejected due to the lack of lensing objects in deep follow-up
images (see Section~\ref{sec:note}). We also carried out additional 
spectroscopy for four objects that did not show a lensing object 
in the images, but which look visually like promising lens candidates. 
We confirmed, however, that they each are a quasar-star pair. 
In conclusion we found four lensed quasars among the color-selected 
candidates. 

We remark that our selection process would reject ``dark lenses'', in 
which the mass-to-light ratio of the lens galaxy is unusually large. The
possibility of dark lenses has been discussed in, e.g., \citet{rusin02} 
and \citet{ryan08}, although there are no unambiguous cases of such
dark lenses in the literature.
 
\subsection{Newly Discovered Binary Quasars}\label{sec:note}
 
We found 29 objects that have multiple quasar components among the 136
candidates. Ten of them are lensed quasars, and four of them are known
pairs of quasars with different redshifts reported in \citet{hennawi06a}. 
Eight systems are pairs of quasars, which we found to have different
redshifts. The remaining seven systems consist of quasar pairs with
identical redshifts. We, however, do not consider them to be lenses,
for the reasons we discuss in what follows. 

{\bf SDSS~J101211.29+365030.7:}
We detect \ion{C}{4} (${\sim}4130${\AA}) and \ion{C}{3]}
(${\sim}5100${\AA}) emission lines at $z=1.678$ in both of the 
components separated by $16\farcs6$, in follow-up spectroscopy at
the ARC 3.5-m telescope. Deep optical and near-infrared
imaging with a detection limit (S/N$\simeq10$) corresponding to 
$M_I{\sim}-21.3$ at $z=1.678$, however, do not show a galaxy
cluster between the two quasar images, which would be necessary to
produce this large image separation. Therefore, this object,
SDSS~J1012+3650, is taken to be a binary quasar.  

{\bf SDSS~J151109.85+335701.7:} 
The images of this candidate show two point sources separated by 
$\theta=1\farcs1$. We detected a signature of a possible
lens galaxy when we subtract two PSFs from the UH88 $I$-band image.
We then obtained spectra of the two stellar components and a deeper 
$I$-band image using the Subaru telescope under good seeing conditions
(0\farcs6). The \ion{C}{2} and \ion{Mg}{2} emission lines redshifted 
to $z=0.799$ have similar shapes in the two objects. The slopes of 
the continua are also similar. We do not find, however, a galaxy 
between the two quasars at a detection limit (S/N$\simeq10$) of 
$M_I{\sim}-19.9$ (at $z=0.799$) in the Subaru image. This suggests 
that the residual flux arises from the host galaxies of the two
quasars. 

{\bf SDSS~J151823.05+295925.4:}
We obtained spectra of the two components using the ARC 3.5-m
telescope. The shapes of the \ion{C}{3]} and \ion{Mg}{2} emission
lines at $z=1.249$ are similar. However, we do not find any galaxies
between the two quasars in a deep optical image taken at UH88
to a detection limit (S/N$\simeq10$) of $M_i{\sim}-20.6$ at $z=1.249$. 
Based on the lensing criteria described in Section~\ref{sec:followup}, 
we conclude that SDSS~J1518+2959 is a binary quasar with
$\theta=5\farcs3$.  

{\bf SDSS~J155218.09+045635.2:}
The spectra of the two components were obtained at the ARC 3.5-m
telescope. While the brighter component has a clear Broad Absorption
Line (BAL) feature in its \ion{C}{4} emission line at $z=1.567$, the
fainter one does not show a BAL feature. In addition, we do not detect
any signature of a massive galaxy cluster in the deep UH88 $I$-band
image, which would be necessary to account for the large image separation 
of $\theta=11\farcs7$.  

{\bf SDSS~J155225.62+300902.0:}
This system was considered to be a promising morphologically-selected 
candidate, with a possible lens galaxy
detected in UH88 $I$-band and  UKIRT $K$-band images. We took spectra
of the two stellar components ($\theta=1\farcs3$) using the Subaru
telescope.  Although the \ion{Mg}{2} emission line of the fainter
component at $z=0.750$ is slightly broader than that of the brighter
component, the two quasar have similar continua. As in the case of
SDSS~J1511+3357, however, we find no galaxy between the two components
at a detection limit  (S/N$\simeq10$) of $M_I{\sim}-19.8$ at
$z=0.750$ in the deep $I$-band Subaru image (seeing ${\sim}0\farcs6$), 
except for extended residuals around the two quasars that represent 
the host galaxies of the quasars. 

{\bf SDSS~J160602.81+290048.7:}
The two stellar components separated by $\theta=3\farcs5$ have similar 
shapes for the \ion{Mg}{2} emission lines at $z=0.770$ and the continua
in our follow-up spectra taken at the ARC 3.5-m telescope. We do not  
find, however, a lens galaxy between the two quasar components at a 
detection limit (S/N$\simeq10$) of $M_i{\sim}-19.3$ at $z=0.770$ in the 
deep $i$-band image with the ARC 3.5-m telescope.  

{\bf SDSS~J163520.04+205225.1:}
We obtained spectra of the two components ($\theta=13\farcs6$) at the
TNG 3.6m telescope. The \ion{C}{4} emission lines at $z=1.775$
revealed that the fainter component is probably a BAL quasar whereas
the brighter component is not. Along with the absence of a lens
cluster of galaxies in deep images taken at the UH88 and KPNO 2.1m
telescopes, we conclude that this object is a binary quasar.  

To summarize, two (SDSS~J1552+0456 and SDSS~J1635+2052) of the seven
candidates are found to be binary quasars from their different
spectral energy distributions.  The other five objects are also 
classified as binary quasars, based on the failure to detect lensing 
objects in our imaging follow-up observations, which were deep enough 
to detect galaxies down to magnitudes significantly fainter than those 
of lensing galaxies for our confirmed lens sample (see 
Figure~\ref{fig:l-angle}).  

\section{IDENTIFIED LENSED QUASARS}\label{sec:dr5_lens}

\subsection{Statistical Sample}\label{sec:statsample}

We construct a DR5 complete sample of lensed quasars with image
separations of $1''<\theta<20''$ and absolute $i$-band (or $I$-band)
magnitude differences less than 1.25~mag for doubles. We do not set
any magnitude difference limits for quadruples. Simulations described
in Paper I suggest that our lens selection is almost complete within
these ranges. From the 13,636 quasars we selected 136 candidates for
lensing, among which  we confirmed ten lensed quasars, six based on 
morphological selection (including the known one, PG1115+080) and four
based on color selection. Eight of the ten meet the criteria we set
for the separation angle and the flux ratio, while
SDSS~J132236.41+105239.4 and SDSS~J134929.84+122706.9 lie outside the
criteria. Including the 11 lensed quasars from the DR3 sample (Paper
II), our DR5 statistical sample consists of 19 lensed quasars selected 
from a sample of 36,287 quasars, as summarized in Table~\ref{tab:lens_dr5stat}.  
The details of the 19 lensed quasars are given in the discovery
papers cited in Table~\ref{tab:lens_dr5stat}. We note that we recover 
all previously known lenses (CASTLES webpage\footnote{
 http://cfa-www.harvard.edu/castles/.}) that 
satisfy our criteria in the area of the DR5 spectroscopic survey.

\begin{figure}
\epsscale{.7}
\plotone{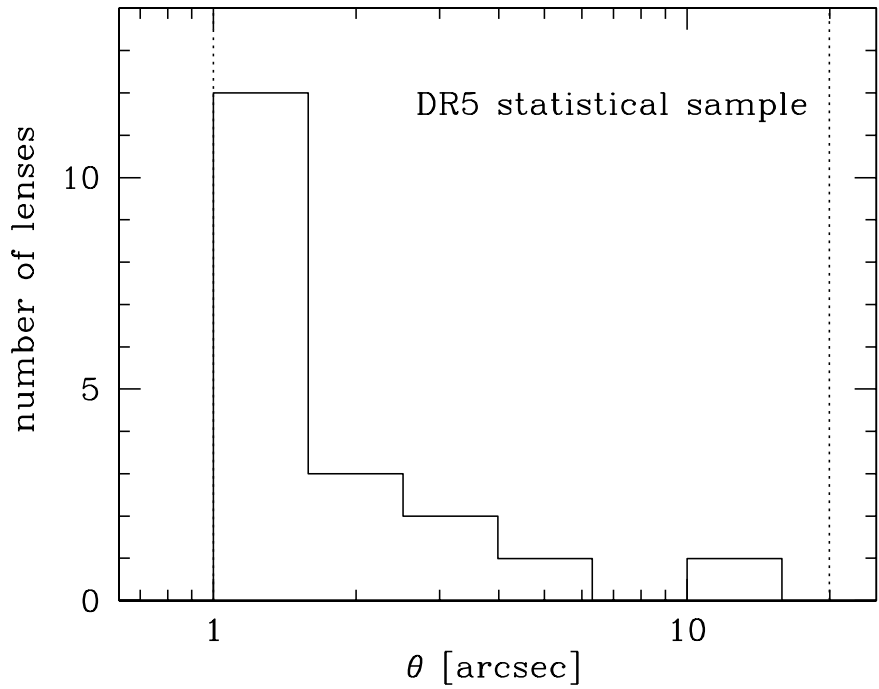}
\caption{The image separation distribution of the SQLS DR5 statistical
lens sample, in bins of $\Delta\log\theta=0.2$. Individual lensed
quasars are listed in Table~\ref{tab:lens_dr5stat}. The dotted lines
indicate upper and lower limits of the image separation for our
statistical lens sample.  
\label{fig:dr5lens}}
\end{figure}

Figure~\ref{fig:dr5lens} shows the distribution of the image separations  
of the statistical sample. We obtain updated constraints on cosmological  
parameters by simply repeating the calculation done in Paper III,
assuming a flat universe. The details of the calculation are given in
Paper III; in brief, we compute the expected numbers of
small-separation lensed quasars for different cosmological models, and
compare them with our statistical lens sample in the image separation
range of $1''<\theta<3''$.  We consider lensing by single elliptical
galaxies which are modeled by singular isothermal ellipsoids. We adopt
the velocity function of \citet{choi07}. The magnification bias is
estimated (see Equations 6, 7, and 8 of Paper III) using the image
separation-dependent magnification factor derived in Paper I and the
quasar luminosity function obtained by \citet{richards05}. As in Paper
III, we require that the PSF magnitude of the quasar  components must
be brighter than the lens galaxy. Since our calculation takes only
early-type galaxies into account, we remove SDSS~J1313+5151 whose lens
galaxy is fit by a S\'{e}rsic profile with $n=1$ \citep{ofek07} and
looks somewhat bluer, and hence is likely to be a late-type
galaxy. From the subsample of 14 lenses, the cosmological constant is
constrained to be  $\Omega_\Lambda=0.84^{+0.06}_{-0.08}({\rm
  stat.})^{+0.09}_{-0.07}({\rm syst.})$. A hypothetical case that 
we have 15 lenses in the subsample (we add one more lens with 
$\theta{\sim}{1\farcs0}$) increases the value of 
$\Omega_\Lambda$ in ${\sim}$0.02. The largest source of 
systematic errors is the uncertainties in the velocity
function of the lens galaxies and its redshift evolution; see Table~2
and Section~4.3 of Paper III for comprehensive discussions of the
systematic errors. If we account for the statistical error only, our
result rejects $\Omega_\Lambda=0$ at the 5$\sigma$ level, as 
estimated from the full likelihood distribution. Next we allow the
dark energy equation of state to vary, and derive constraints in the
two-dimensional parameter space of $\Omega_M$ and $w$. By combining
our result with the results from the SDSS baryon acoustic oscillation
measurements in the SDSS galaxy two-point correlation function
\citep{eisenstein05}, we obtain 
$\Omega_M=0.23^{+0.04}_{-0.03}({\rm stat.})^{+0.03}_{-0.04}({\rm
  syst.})$ and $w=-1.4 \pm 0.3({\rm stat.})^{+0.3}_{-0.4}({\rm syst.})$.

The fraction of quadruple lenses in the DR5 statistical sample is 16\%
(see Table~\ref{tab:lens_dr5stat}), which is lower than the fraction
in the CLASS survey of 46\%. This might be ascribed to the shallower
slope of the luminosity  function at the survey flux limit of SQLS
than that at the CLASS limit \citep[see, e.g.,][for more detailed
discussion]{oguri07b}.   

\subsection{Additional Lensed Quasars}\label{sec:addlens}

We also searched for lensed quasars in the DR5 sample which do not
satisfy the conditions, $0.6 < z < 2.2$, $i<19.1$ and {\tt
  PSF\_WIDTH}$<1\farcs8$, with the understanding that the resulting
sample will not be complete. The candidate selection is somewhat
extended from that used to make the statistical sample. For high
redshift  ($z>2.2$) quasars, for instance, we use the {\tt star\_L}
criteria for the $griz$ bands rather than the $ugri$ bands
\citep[e.g.,][]{inada09}. We discovered two lensed quasars,
SDSS~J0819+5356 \citep{inada09} at $z_s=2.24$ and SDSS~J2343$-$0050
\citep[ULAS~J234311.93-005034.0;][]{jackson08} with $i=20.10$ at
$z_s=0.787$. The second of these lensed quasars was previously 
discovered by \citet{jackson09} from the UKIDSS \citep[UKIRT Infrared 
Deep Sky Survey;][]{lawrence07} and SDSS. There are two lenses among 
the DR5 quasars known from other surveys, SDSS~J1004+1229 (J1004+1229; 
see CASTLES webpage) and  
SDSS~J0820+0812 \citep[ULAS~J082016.1+081216;][]{jackson08}. They 
are not identified in our selection algorithm because of the large
magnitude differences $|{\Delta} I|>1.25$ between the two images.
 Together with 11 lenses that are derived from the DR3 sample but do
 not satisfy our selection criteria \citep{surdej87,bade97,morgan01,
reimers02,winn02,johnston03,morgan03,pindor04,inada07,inada08,oguri08b}, 
we present a list of 17 additional lensed quasars in the entire
DR5 quasar sample in Table~\ref{tab:lens_dr5add}. We note that SDSS~J1322+1052
and SDSS~J1349+1227 (Section~\ref{sec:statsample}) are included in this
additional sample. The details of all lensed quasars in the additional
sample are also given in the references cited in Table~\ref{tab:lens_dr5add}. 

\section{SUMMARY}\label{sec:summary}

We have completed our systematic lensed quasar search in the SDSS-I
quasar sample presented in the DR5 quasar catalog \citep{schneider07}. 
With follow-up observations for 136 lens candidates, we found ten lensed 
quasars beyond our DR3 sample. Eight of them, including 
one previously known, are catalogued in our complete statistical lensed
quasar sample within specified separation ranges and magnitude
differences. These conditions minimize our selection bias for lensed
systems.  Combining with the result from DR3, we present a complete
sample of 19 lensed quasars selected from 36,278 quasars, where
three of them have quadruple-image configurations. We then updated
cosmological constraints obtained in Paper III, and obtained 
results consistent with other cosmological observations
\citep[e.g.,][]{komatsu09}. 

In addition to the 19 lensed quasars in our complete sample, the DR5
quasar catalog contains at least 17 additional lenses. Two were
discovered among the 136 lens candidates considered here but excluded 
from our complete lens sample, and two were discovered from quasars 
outside the source sample of the 36,278 quasars. The remaining 13 
lenses include two previously known lenses and 11 lenses from the DR3 
sample (Paper II). These numbers may be compared with those of the CLASS, 
which contains 13 lenses in their statistical sample with well defined 
criteria and 22 lenses in total among 8958 quasars. 

In this paper, we have also report the discovery of 7 binary quasars 
with nearly identical redshifts, as well as 8 projected quasar
pairs. These
quasar pairs are a useful addition to the studies of the small-scale
correlation function and interaction of quasars
\citep{hennawi06a,myers08,green10} and the spatial distribution of
absorbers \citep{bowen06,hennawi06b,tytler09}.  

The quasar sample used in the present work is constructed from the full
SDSS-I data set, and hence the work represents the completion of our
statistical lens sample in the SDSS-I. We plan to continue our lens
survey further to construct the SQLS lens sample from the SDSS-II,
using the DR7 quasar catalog \citep{schneider10}. More detailed
analysis of cosmological constraints will be presented elsewhere.

\acknowledgments

N.~I. acknowledges support from the Special Postdoctoral Researcher 
Program of RIKEN, the RIKEN DRI Research Grant, and MEXT KAKENHI 21740151.
This work was supported in part by Department of Energy contract
DE-AC02-76SF00515.
I.~K. acknowledges support by Grant-in-Aid for JSPS Fellows and
Grant-in-Aid for Scientific Research on Priority Areas No. 467.
M-S.~S. and M.~A.~S. acknowledge the support of NSF grant AST-0707266. 
A.~C. is supported by grants from MIDEPLAN (ICM/P06-045-F) and CONICYT 
(FONDAP 15010003 and PFB 06). J.~R.~G acknowledges the support of NSF 
grant AST-0406713. C.~S.~K. is supported by NSF grant AST-0708082.

Use of the UH 2.2-m telescope and the UKIRT 3.8-m telescope for the
observations is supported by the National Astronomical
Observatory of Japan (NAOJ). 
This work is also based in part on observations obtained with the MDM
2.4m Hiltner telescope, which is owned and
operated by a consortium consisting of Columbia University, Dartmouth
College, the University of Michigan, the Ohio State University and
Ohio University. 
Some of the data presented herein were obtained at the W.M. Keck
Observatory, which is operated as a scientific partnership among the
California Institute of Technology, the University of California and the
National Aeronautics and Space Administration. The Keck Observatory was 
made possible by the generous financial support of the W.M. Keck
Foundation. 
Based in part on observations made with a telescope (ESO 3.6-m) 
at the European Southern Observatories La Silla in Chile.
Based in part on observations obtained with the Apache Point Observatory
3.5-meter telescope, which is owned and operated by the Astrophysical
Research Consortium. 
Based on data collected at Subaru Telescope, which is operated by the 
National Astronomical Observatory of Japan, and obtained from the SMOKA, 
which is operated by the Astronomy Data Center, National Astronomical 
Observatory of Japan.
The WB 6.5-m telescope is the first telescope of
the Magellan Project; a collaboration between the Observatories of
the Carnegie Institution of Washington, University of
Arizona, Harvard University, University of Michigan, and
Massachusetts Institute of Technology to construct two 6.5 Meter
optical telescopes in the southern hemisphere.
The WIYN Observatory is a joint facility of the University of
Wisconsin-Madison, Indiana University, Yale University, and the
National Optical Astronomy Observatories. 
Telescopio Nazionale Galileo (TNG) operated on the island of La Palma by the 
Fundacion Galileo Galilei of the INAF (Istituto Nazionale di Astrofisica) at 
the Spanish Observatorio del Roque de los Muchachos of the Instituto de 
Astrofisica de Canarias.
Based in part on observations with Kitt Peak National Observatory, which is 
operated by AURA under cooperative agreement with the National Science 
Foundation. 

Funding for the SDSS and SDSS-II was provided by the Alfred P. Sloan 
Foundation, the Participating Institutions, the National Science 
Foundation, the U.S. Department of Energy, the National Aeronautics and 
Space Administration, the Japanese Monbukagakusho, the Max Planck Society, 
and the Higher Education Funding Council for England.

The SDSS was managed by the Astrophysical Research Consortium for the 
Participating Institutions. The Participating Institutions were the 
American Museum of Natural History, Astrophysical Institute Potsdam, 
University of Basel, Cambridge University, Case Western Reserve University, 
University of Chicago, Drexel University, Fermilab, the Institute for 
Advanced Study, the Japan Participation Group, Johns Hopkins University, 
the Joint Institute for Nuclear Astrophysics, the Kavli Institute for 
Particle Astrophysics and Cosmology, the Korean Scientist Group, the Chinese 
Academy of Sciences (LAMOST), Los Alamos National Laboratory, the 
Max-Planck-Institute for Astronomy (MPIA), the Max-Planck-Institute for 
Astrophysics (MPA), New Mexico State University, Ohio State University, 
University of Pittsburgh, University of Portsmouth, Princeton University, 
the United States Naval Observatory, and the University of Washington.

\clearpage

\begin{deluxetable}{lr}
\tabletypesize{\footnotesize}
\tablecaption{NUMBERS OF CANDIDATES\label{tab:cand}}
\tablewidth{0pt}
\tablehead{ 
\colhead{} & 
\colhead{Number} }
\startdata
\\
SDSS DR5 spectroscopically confirmed quasars & 77,429 \\
DR5 quasars {\it not} passing the criteria ($0.6<z<2.2$, $i<19.1$, and 
{\tt PSF\_WIDTH}$<1\farcs8$) & $-$41,142 \\ 
\\
Source quasar sample                                           & 36,287 \\ 
DR3 quasars (checked in Paper II)                                 &
$-$22,651\tablenotemark{a} \\
\\
Source quasars to complete the DR5 statistical lens sample        & 13,636 \\
\\
\hline
\\
Initial morphologically selected candidates (using {\tt star\_L} in $ugri$)  &   
401  \\  
Rejected by GALFIT fitting                                        & $-$348  \\
Rejected by visual inspection                                     &   $-$3  \\
\\ 
Final morphologically selected  candidates for follow-up          &     50  \\   
\\
\hline
\\
Initial color selected candidates                                 &    140  \\  
Rejected by FIRST image check                                     &   $-$5  \\
Rejected by visual inspection                                     &   $-$6  \\
Rejected by searching for possible lens objects                   &  $-$41  \\
\\
Final color selected candidates for follow-up                     &     88  \\   
\\
\hline
\\
Final total (morphological+color) candidates for follow-up  & 136\tablenotemark{b}  \\
\\
\enddata

\tablenotetext{a}{The number of the source quasars in the paper II is 22,683, 
but 32 DR3 quasars do not meet the criteria in the DR5 catalog.}
\tablenotetext{b}{Two candidates are selected by both the morphological 
and color selection algorithms. They are listed in Table~\ref{tab:can_mor} as 
morphologically selected candidates.}
 
\end{deluxetable}

\clearpage

\begin{deluxetable}{lccccllll}
\tabletypesize{\footnotesize}
\rotate
\tablecaption{Morphologically Selected Candidates\label{tab:can_mor}}
\tablewidth{0pt}
\tablehead{ 
\colhead{Object} & 
\colhead{redshift\tablenotemark{a}} &
\colhead{$i$\tablenotemark{b}} & 
\colhead{$\theta_{\rm SDSS}$\tablenotemark{c}} & 
\colhead{$|{\Delta} i|$\tablenotemark{c}} & 
\colhead{image\tablenotemark{d}} & 
\colhead{spec\tablenotemark{d}} & 
\colhead{comment} & 
\colhead{Ref.} 
}
\startdata
SDSS~J024519.65$-$005113.0  &  1.545  &  18.78  &  1.12  &  1.24  &  UF($K$)        
                    &  \nodata\phn  &  QSO+galaxy           &  \nodata\phn  \\ 
SDSS~J030556.81+005701.7    &  0.893  &  19.00  &  2.09  &  1.46  &  UF($K$)        
                    &  \nodata\phn  &  QSO+galaxy           &  \nodata\phn  \\ 
SDSS~J074653.03+440351.3    &  1.998  &  18.71  &  1.07  &  0.04  & 
8k($VRI$),RE($r$),Op($I$)           &  ES           &  SDSS lens            &  1    
       \\ 
SDSS~J074942.51+171512.1    &  2.163  &  18.87  &  1.01  &  0.82  &  Te($I$)        
                    &  \nodata\phn  &  no lens object    &  \nodata\phn  \\ 
SDSS~J080009.98+165509.4    &  0.708  &  17.97  &  0.79  &  0.03  &  Te($I$)        
                    &  \nodata\phn  &  no lens object    &  \nodata\phn  \\ 
SDSS~J080623.70+200631.8    &  1.537  &  18.88  &  1.42  &  0.26  & 
8k($VRI$),QU($H$),NR($K'$)          &  ES           &  SDSS lens            &  2    
       \\ 
SDSS~J082312.13+264415.7    &  1.855  &  18.16  &  1.40  &  1.42  &  Te($I$)        
                    &  \nodata\phn  &  QSO+galaxy           &  \nodata\phn  \\ 
SDSS~J082341.08+241805.4    &  1.811  &  16.90  &  0.58  &  0.04  &  8k($V$)        
                    &  \nodata\phn  &  single QSO           &  \nodata\phn  \\ 
SDSS~J083240.71+060759.3    &  0.808  &  19.04  &  1.58  &  1.92  &  Te($I$)        
                    &  \nodata\phn  &  no lens object    &  \nodata\phn  \\ 
SDSS~J094713.15+024743.6    &  0.641  &  18.99  &  1.57  &  1.11  &  Te($I$)        
                    &  \nodata\phn  &  no lens object    &  \nodata\phn  \\ 
SDSS~J095237.05+290834.6    &  1.413  &  18.22  &  4.55  &  1.14  &  Te($I$)        
                    &  \nodata\phn  &  QSO+galaxy           &  \nodata\phn  \\ 
SDSS~J104901.21+121214.1    &  1.572  &  19.09  &  1.50  &  0.84  &  \nodata\phn    
                    &  EF           &  QSO+unknown(not QSO) &  \nodata\phn  \\ 
SDSS~J105545.44+462839.5    &  1.249  &  18.75  &  1.12  &  0.89  & 
8k($VRI$),FO($RI$),Te($RI$),NF($H$) &  FO           &  SDSS lens            &  3    
       \\ 
SDSS~J110456.56+130711.1    &  1.778  &  18.47  &  1.07  &  1.11  &  Te($I$)        
                    &  FO           &  QSO+star             &  \nodata\phn  \\ 
SDSS~J111816.95+074558.1    &  1.736  &  15.96  &  2.26  &  1.92  &  \nodata\phn    
                    &  \nodata\phn  &  known lens (PG1115)  &  4            \\ 
SDSS~J112508.26+303141.3    &  1.960  &  17.62  &  0.43  &  1.17  &  Te($I$)        
                    &  \nodata\phn  &  single QSO           &  \nodata\phn  \\ 
SDSS~J113831.39+151215.3    &  0.659  &  18.42  &  0.50  &  2.06  &  Te($I$)        
                    &  \nodata\phn  &  single QSO           &  \nodata\phn  \\ 
SDSS~J114217.47+451447.7    &  1.782  &  18.95  &  3.93  &  0.26  &  \nodata\phn    
                    &  DA           &  QSO+star             &  \nodata\phn  \\ 
SDSS~J114443.69+350539.6    &  0.604  &  18.74  &  0.79  &  1.86  &  Te($I$)        
                    &  \nodata\phn  &  single QSO           &  \nodata\phn  \\ 
SDSS~J115619.49+460313.8    &  1.235  &  18.11  &  0.44  &  1.75  &  Te($I$)        
                    &  \nodata\phn  &  single QSO           &  \nodata\phn  \\ 
SDSS~J115800.91+120439.4    &  1.615  &  18.57  &  1.84  &  1.24  &  Te($I$)        
                    &  \nodata\phn  &  QSO+galaxy           &  \nodata\phn  \\ 
SDSS~J120118.92+401318.1    &  1.933  &  18.56  &  1.54  &  1.51  &  Te($I$)        
                    &  \nodata\phn  &  QSO+galaxy           &  \nodata\phn  \\ 
SDSS~J120239.66+455429.7    &  1.071  &  18.46  &  0.40  &  0.94  &  Te($I$)        
                    &  \nodata\phn  &  single QSO           &  \nodata\phn  \\ 
SDSS~J120348.93+325542.5    &  1.217  &  18.89  &  0.47  &  1.80  &  Te($I$)        
                    &  \nodata\phn  &  single QSO           &  \nodata\phn  \\ 
SDSS~J120730.01+125057.6    &  0.752  &  18.20  &  0.54  &  1.08  &  Te($I$)        
                    &  \nodata\phn  &  single QSO           &  \nodata\phn  \\ 
SDSS~J120912.45+143602.3    &  1.497  &  18.10  &  0.42  &  0.48  &  Te($I$)        
                    &  \nodata\phn  &  single QSO           &  \nodata\phn  \\ 
SDSS~J121357.15+083202.2    &  0.811  &  18.01  &  0.41  &  0.87  &  Te($I$)        
                    &  \nodata\phn  &  single QSO           &  \nodata\phn  \\
SDSS~J122848.03+151018.4    &  1.118  &  17.59  &  0.40  &  1.38  &  Te($I$)        
                    &  \nodata\phn  &  single QSO           &  \nodata\phn  \\
SDSS~J130225.24+332933.2    &  0.922  &  18.19  &  0.50  &  1.34  &  Te($I$)        
                    &  \nodata\phn  &  single QSO           &  \nodata\phn  \\
SDSS~J132236.41+105239.4    &  1.716  &  18.24  &  1.88  &  1.38  & 
8k($V$),Te($VRI$),NF($H$)           &  WF           &  SDSS lens            &  5    
       \\ 
SDSS~J134322.04+314827.7    &  0.930  &  18.32  &  0.49  &  0.33  &  Te($I$)        
                    &  \nodata\phn  &  single QSO           &  \nodata\phn  \\
SDSS~J135143.59+245248.8    &  1.286  &  19.06  &  1.47  &  1.34  &  Te($I$)        
                    &  \nodata\phn  &  no lens object    &  \nodata\phn  \\
SDSS~J135306.34+113804.7    &  1.623  &  16.48  &  1.33  &  0.97  & 
8k($VRI$),QU($H$),Ma($gi$)          &  ES           &  SDSS lens            &  2    
       \\
SDSS~J135404.14+110725.7    &  1.318  &  18.26  &  1.34  &  1.31  &  Te($I$)        
                    &  \nodata\phn  &  no lens object    &  \nodata\phn  \\ 
SDSS~J141202.70+354247.2    &  1.360  &  19.02  &  1.00  &  1.42  &  Te($I$)        
                    &  \nodata\phn  &  QSO+galaxy           &  \nodata\phn  \\
SDSS~J141910.20+420746.9    &  0.874  &  17.04  &  0.45  &  1.38  &  Te($I$)        
                    &  \nodata\phn  &  single QSO           &  \nodata\phn  \\
SDSS~J142030.50+353328.6    &  1.689  &  18.79  &  1.35  &  2.06  &  8k($I$)        
                    &  \nodata\phn  &  QSO+galaxy           &  \nodata\phn  \\
SDSS~J142326.96+093216.5    &  0.633  &  19.03  &  1.58  &  1.75  &  Te($I$)        
                    &  \nodata\phn  &  no lens object    &  \nodata\phn  \\
SDSS~J143344.39+113941.9    &  1.447  &  18.42  &  0.73  &  1.31  &  Te($V$)        
                    &  \nodata\phn  &  single QSO           &  \nodata\phn  \\
SDSS~J143452.44+133459.5    &  1.689  &  18.26  &  0.55  &  1.34  &  Te($I$)        
                    &  \nodata\phn  &  single QSO           &  \nodata\phn  \\
SDSS~J144618.91+114446.2    &  1.243  &  17.66  &  0.43  &  0.65  &  Te($I$)        
                    &  \nodata\phn  &  single QSO           &  \nodata\phn  \\
SDSS~J145307.06+331950.5    &  1.192  &  18.89  &  1.69  &  1.17  &  \nodata\phn    
                    &  ES           &  QSO+star             &  \nodata\phn  \\ 
SDSS~J150020.24+340038.9    &  0.732  &  18.90  &  0.63  &  1.51  &  Te($I$)        
                    &  \nodata\phn  &  single QSO           &  \nodata\phn  \\
SDSS~J151109.85+335701.7    &  0.799  &  18.96  &  1.10  &  0.69  & 
8k($V$),Te($I$),FO($I$)             &  FO           &  binary QSO ($z=0.799,0.799$) 
&  \nodata\phn  \\ 
SDSS~J152938.10+300351.1    &  0.641  &  18.36  &  0.44  &  1.99  &  Te($I$)        
                    &  \nodata\phn  &  single QSO           &  \nodata\phn  \\
SDSS~J153325.42+361915.5    &  0.681  &  18.90  &  0.42  &  0.87  &  Te($I$)        
                    &  \nodata\phn  &  single QSO           &  \nodata\phn  \\
SDSS~J155000.01+300223.6    &  0.657  &  18.68  &  0.45  &  0.48  &  Te($V$)        
                    &  \nodata\phn  &  single QSO           &  \nodata\phn  \\
SDSS~J155225.62+300902.0    &  0.750  &  18.85  &  1.25  &  0.57  & 
8k($VI$),Op($I$),FO($I$),UF($K$)    &  FO           &  binary QSO ($z=0.752,0.752$) 
&  \nodata\phn  \\ 
SDSS~J165743.05+221149.1    &  1.780  &  17.88  &  1.17  &  1.20  &  MM($z$),UF($K$)
                    &  \nodata\phn  &  QSO+galaxy           &  \nodata\phn  \\ 
SDSS~J221227.74+005140.5    &  1.773  &  18.88  &  1.76  &  1.55  &  UF($K$)        
                    &  \nodata\phn  &  no lens object    &  \nodata\phn  \\ 
\enddata
\tablenotetext{a}{Redshifts from the SDSS DR5 quasar catalog.} 
\tablenotetext{b}{$i$-band PSF magnitudes with Galactic extinction
   corrections from the SDSS DR5 quasar catalog.}  
\tablenotetext{c}{Image separations ($\theta_{\rm SDSS}$) in units of
   arcsec and absolute $i$-band magnitude differences ($|{\Delta} i|$) between the
expected
   two components, derived from fitting the SDSS $i$-band image of each candidate 
   with two PSFs using GALFIT.} 
\tablenotetext{d}{Instruments (and filters) used for the follow-up observations. 
   UF: UFTI at UKIRT, 
   8k: UH8k at UH88, Te: Tek2k CCD at UH88, Op: Optic CCD at UH88, QU: QUIRC at
UH88, WF: WFGS2 at UH88,
   RE: RETROCAM at MDM 2.4m, 
   ES: ESI at Keck, NR: NIRC at Keck,
   EF: EFOSC2 at ESO 3.6m,
   NF: NICFPS at ARC 3.5m, DA: DIS III at ARC 3.5m,
   FO: FOCAS at Subaru, 
   Ma: MagIC at WB 6.5m,
   MM: MiniMo at WIYN.
   }
\tablerefs{
   (1) \citealt{inada07}; 
   (2) \citealt{inada06};
   (3) \citealt{kayo09};
   (4) \citealt{weymann80}; 
   (5) \citealt{oguri08b}.
   }
\end{deluxetable}

\clearpage

\begin{deluxetable}{lccrllll}
\tabletypesize{\footnotesize}
\rotate
\tablecaption{Color Selected Candidates\label{tab:can_col}}
\tablewidth{0pt}
\tablehead{ 
\colhead{Object} & 
\colhead{redshift\tablenotemark{a}} &
\colhead{$i$\tablenotemark{b}} & 
\colhead{$\theta_{\rm SDSS}$\tablenotemark{c}} & 
\colhead{image\tablenotemark{d}} & 
\colhead{spec\tablenotemark{d}} & 
\colhead{comment} & 
\colhead{Ref.} 
}
\startdata
SDSS~J004018.21+005530.9   &  2.019        &  18.62  &         &                    
&               &                               &               \\ 
SDSS~J004018.68+005525.9   &  2.080        &  19.13  &   8.53  &  \nodata\phn       
&  DA           &  QSO pair                     &  1  \\ 
\hline
SDSS~J014917.10$-$002141.6 &  1.688        &  18.35  &         &                    
&               &                               &               \\ 
SDSS~J014917.47$-$002158.5 &  2.160        &  19.46  &  17.67  &  \nodata\phn       
&  DA           &  QSO pair                     &  1  \\ 
\hline
SDSS~J032029.75+000650.0   &  1.704        &  19.05  &         &                    
&               &                               &               \\ 
SDSS~J032030.90+000658.0   &  \nodata\phn  &  20.34  &  18.99  &  UF($K$)           
&  \nodata\phn  &  no lens object            &  \nodata\phn  \\ 
\hline
SDSS~J034347.00$-$000706.5 &  1.975        &  18.85  &         &                    
&               &                               &               \\ 
SDSS~J034347.48$-$000658.9 &  \nodata\phn  &  19.30  &  10.49  &  UF($K$)           
&  \nodata\phn  &  no lens object            &  \nodata\phn  \\ 
\hline
SDSS~J072653.68+394706.9   &  1.599        &  18.92  &         &                    
&               &                               &               \\ 
SDSS~J072653.66+394710.6   &  \nodata\phn  &  18.84  &   3.69  &  Te($I$)           
&  \nodata\phn  &  no lens object            &  \nodata\phn  \\ 
\hline
SDSS~J074550.99+503423.1   &  1.737        &  18.98  &         &                    
&               &                               &               \\ 
SDSS~J074550.78+503430.3   &  \nodata\phn  &  19.84  &   7.44  &  Te($I$)           
&  \nodata\phn  &  no lens object            &  \nodata\phn  \\ 
\hline
SDSS~J075403.19+193740.9   &  1.540        &  19.05  &         &                    
&               &                               &               \\ 
SDSS~J075403.60+193734.2   &  \nodata\phn  &  19.98  &   8.87  &  \nodata\phn       
&  DA           &  QSO+star                     &  \nodata\phn  \\ 
\hline
SDSS~J080932.70+193847.2   &  1.670        &  18.49  &         &                    
&               &                               &               \\ 
SDSS~J080931.82+193849.3   &  \nodata\phn  &  19.28  &  12.68  &  Te($I$)           
&  \nodata\phn  &  no lens object            &  \nodata\phn  \\ 
\hline
SDSS~J081210.96+070826.2   &  1.862        &  17.65  &         &                    
&               &                               &               \\ 
SDSS~J081210.93+070838.7   &  \nodata\phn  &  16.89  &  12.39  &  Te($I$)           
&  \nodata\phn  &  no lens object            &  \nodata\phn  \\ 
\hline
SDSS~J083228.49+563234.2   &  0.683        &  18.94  &         &                    
&               &                               &               \\ 
SDSS~J083228.53+563237.3   &  \nodata\phn  &  18.91  &   3.09  &  Te($I$)           
&  DA           &  QSO+star                     &  \nodata\phn  \\ 
\hline
SDSS~J084109.72+250200.2   &  1.227        &  19.06  &         &                    
&               &                               &               \\ 
SDSS~J084109.88+250216.0   &  \nodata\phn  &  20.13  &  15.89  &  Te($I$)           
&  \nodata\phn  &  no lens object            &  \nodata\phn  \\ 
\hline
SDSS~J084359.79+073229.7   &  2.175        &  19.02  &         &                    
&               &                               &               \\ 
SDSS~J084359.89+073215.9   &  \nodata\phn  &  20.28  &  13.97  &  Te($I$)           
&  \nodata\phn  &  no lens object            &  \nodata\phn  \\ 
\hline
SDSS~J085705.91+270149.0   &  1.419        &  19.08  &         &                    
&               &                               &               \\ 
SDSS~J085706.14+270147.6   &  \nodata\phn  &  20.06  &   3.36  &  Te($I$)           
&  \nodata\phn  &  no lens object            &  \nodata\phn  \\ 
\hline
SDSS~J090323.94+313445.6   &  1.217        &  18.99  &         &                    
&               &                               &               \\ 
SDSS~J090324.19+313443.6   &  \nodata\phn  &  19.23  &   3.78  &  Te($I$)           
&  \nodata\phn  &  no lens object            &  \nodata\phn  \\ 
\hline
SDSS~J090505.55+085826.4   &  1.082        &  19.03  &         &                    
&               &                               &               \\ 
SDSS~J090505.51+085829.3   &  \nodata\phn  &  20.22  &   2.88  &  Te($I$)           
&  \nodata\phn  &  no lens object            &  \nodata\phn  \\ 
\hline
SDSS~J092722.27+343321.2   &  1.393        &  18.71  &         &                    
&               &                               &               \\ 
SDSS~J092721.58+343309.1   &  \nodata\phn  &  19.99  &  14.87  &  \nodata\phn       
&  DA           &  not QSO                      &  \nodata\phn  \\ 
\hline
SDSS~J093514.07+372140.7   &  1.798        &  18.34  &         &                    
&               &                               &               \\ 
SDSS~J093514.41+372145.1   &  \nodata\phn  &  18.35  &   5.91  &  \nodata\phn       
&  DA           &  QSO+star                     &  \nodata\phn  \\ 
\hline
SDSS~J094115.37+305810.3   &  1.193        &  19.02  &         &                    
&               &                               &               \\ 
SDSS~J094115.49+305808.6   &  \nodata\phn  &  20.24  &   2.39  &  Te($I$)           
&  \nodata\phn  &  QSO+galaxy                   &  \nodata\phn  \\ 
\hline
SDSS~J094132.43+112913.1   &  1.535        &  18.57  &         &                    
&               &                               &               \\ 
SDSS~J094131.42+112917.3   &  \nodata\phn  &  18.45  &  15.51  &  \nodata\phn       
&  DA           &  QSO+star                     &  \nodata\phn  \\ 
\hline
SDSS~J094903.46+280021.9   &  1.563        &  18.96  &         &                    
&               &                               &               \\ 
SDSS~J094903.56+280023.2   &  \nodata\phn  &  19.69  &   1.72  &  UF($K$)           
&  FO           &  QSO+star                     &  \nodata\phn  \\ 
\hline
SDSS~J095126.56+324601.4   &  1.552        &  18.41  &         &                    
&               &                               &               \\ 
SDSS~J095127.53+324549.1   &  1.928        &  19.59  &  17.30  &  \nodata\phn       
&  DA           &  QSO pair                     &  \nodata\phn  \\ 
\hline
SDSS~J095454.99+373419.9   &  1.884        &  18.36  &         &                    
&               &                               &               \\ 
SDSS~J095454.74+373419.8   &  1.540        &  19.19  &   3.14  &  \nodata\phn       
&  DA           &  QSO pair                     &  1  \\
\hline
SDSS~J095820.72+291901.1   &  1.537        &  18.65  &         &                    
&               &                               &               \\ 
SDSS~J095821.77+291855.5   &  \nodata\phn  &  19.60  &  14.75  &  Te($I$)           
&  \nodata\phn  &  no lens object            &  \nodata\phn  \\ 
\hline
SDSS~J101211.29+365030.7   &  1.678        &  18.81  &         &                    
&               &                                 &               \\ 
SDSS~J101211.07+365014.3   &  1.678        &  20.01  &  16.60  &  Te($I$),NF($H$)   
&  DA           &  no lens object, binary QSO  &  \nodata\phn  \\ 
\hline
SDSS~J101753.63+414931.4   &  2.114        &  18.86  &         &                    
&               &                               &               \\ 
SDSS~J101752.33+414922.6   &  \nodata\phn  &  18.97  &  17.08  &  \nodata\phn       
&  DA           &  QSO+star                     &  \nodata\phn  \\
\hline
SDSS~J102335.90+384909.4   &  1.362        &  18.95  &         &                    
&               &                               &               \\ 
SDSS~J102336.00+384908.2   &  \nodata\phn  &  19.13  &   1.71  &  Te($I$)           
&  \nodata\phn  &  no lens object            &  \nodata\phn  \\
\hline
SDSS~J103149.35+371055.2   &  2.116        &  19.06  &         &                    
&               &                               &               \\ 
SDSS~J103150.14+371052.6   &  \nodata\phn  &  17.76  &   9.73  &  \nodata\phn       
&  DA           &  QSO+star                     &  \nodata\phn  \\
\hline
SDSS~J103419.11+124608.0   &  2.104        &  18.77  &         &                    
&               &                               &               \\ 
SDSS~J103419.16+124609.9   &  \nodata\phn  &  18.69  &   1.95  &  Te($I$)           
&  \nodata\phn  &  no lens object            &  \nodata\phn  \\
\hline
SDSS~J104655.40+292045.7   &  1.139        &  18.94  &         &                    
&               &                               &               \\ 
SDSS~J104654.72+292052.3   &  \nodata\phn  &  19.47  &  11.03  &  Te($I$)           
&  \nodata\phn  &  no lens object            &  \nodata\phn  \\
\hline
SDSS~J111727.07+420003.0   &  1.154        &  19.01  &         &                    
&               &                               &               \\ 
SDSS~J111727.08+420000.3   &  \nodata\phn  &  20.25  &   2.78  &  Te($I$)           
&  \nodata\phn  &  no lens object            &  \nodata\phn  \\
\hline
SDSS~J112331.19+134954.5   &  1.352        &  18.99  &         &                    
&               &                               &               \\ 
SDSS~J112331.15+134952.9   &  \nodata\phn  &  19.72  &   1.73  &  Te($I$)           
&  FO           &  QSO+star                     &  \nodata\phn  \\
\hline
SDSS~J112856.75+143352.2   &  1.533        &  18.37  &         &                    
&               &                               &               \\ 
SDSS~J112856.01+143349.4   &  \nodata\phn  &  18.58  &  11.22  &  Te($I$)           
&  \nodata\phn  &  no lens object            &  \nodata\phn  \\
\hline
SDSS~J113358.60+063625.7   &  1.570        &  17.97  &         &                    
&               &                               &               \\ 
SDSS~J113358.01+063630.2   &  \nodata\phn  &  19.02  &   9.90  &  Te($I$)           
&  \nodata\phn  &  no lens object            &  \nodata\phn  \\
\hline
SDSS~J120450.54+442835.8   &  1.142        &  18.79  &         &                    
&               &                               &               \\ 
SDSS~J120450.78+442834.2   &  1.814        &  19.20  &   3.06  &  \nodata\phn       
&  DA           &  QSO pair                     &  1  \\
\hline
SDSS~J120629.64+433217.5   &  1.790        &  18.47  &         &                    
&               &                               &               \\ 
SDSS~J120629.65+433220.6   &  1.790        &  19.16  &   3.04  &  8k($VRI$)         
&  DA           &  SDSS lens                    &  2            \\
\hline
SDSS~J121646.05+352941.5   &  2.012        &  19.08  &         &                    
&               &                               &               \\ 
SDSS~J121645.93+352941.6   &  2.012        &  19.86  &   1.45  &  Te($VRI$),NF($H$) 
&  WF           &  SDSS lens                    &  3            \\
\hline
SDSS~J121653.10+350503.7   &  1.854        &  18.52  &         &                    
&               &                               &               \\ 
SDSS~J121654.53+350510.8   &  1.800        &  19.33  &  18.82  &  \nodata\phn       
&  DA           &  QSO pair                     &  \nodata\phn  \\
\hline
SDSS~J122109.50+364732.2   &  1.859        &  18.92  &         &                    
&               &                               &               \\ 
SDSS~J122109.47+364739.0   &  0.470        &  19.87  &   6.76  &  \nodata\phn       
&  DA           &  QSO pair                     &  \nodata\phn  \\
\hline
SDSS~J122332.63+412939.2   &  0.804        &  18.72  &         &                    
&               &                               &               \\ 
SDSS~J122334.17+412945.6   &  \nodata\phn  &  18.77  &  18.36  &  \nodata\phn       
&  DA           &  not QSO                      &  \nodata\phn  \\ 
\hline
SDSS~J123140.73+395722.9   &  1.572        &  18.98  &         &                    
&               &                               &               \\ 
SDSS~J123139.73+395721.7   &  \nodata\phn  &  19.56  &  11.63  &  \nodata\phn       
&  DA           &  QSO+star                     &  \nodata\phn  \\
\hline
SDSS~J123449.46+084320.2   &  1.830        &  18.86  &         &                    
&               &                               &               \\ 
SDSS~J123448.88+084303.9   &  \nodata\phn  &  19.51  &  18.49  &  \nodata\phn       
&  DA           &  QSO+star                     &  \nodata\phn  \\
\hline
SDSS~J123823.37+463439.1   &  1.411        &  19.08  &         &                    
&               &                               &               \\ 
SDSS~J123821.84+463445.8   &  1.457        &  19.97  &  17.15  &  \nodata\phn       
&  DA           &  QSO pair                     &  \nodata\phn  \\
\hline
SDSS~J125104.28+151144.9   &  2.075        &  18.60  &         &                    
&               &                               &               \\ 
SDSS~J125103.79+151159.8   &  \nodata\phn  &  19.24  &  16.52  &  \nodata\phn       
&  DA           &  not QSO                      &  \nodata\phn  \\
\hline
SDSS~J131019.76+401724.2   &  2.054        &  18.81  &         &                    
&               &                               &               \\ 
SDSS~J131019.52+401713.9   &  \nodata\phn  &  20.10  &  10.75  &  \nodata\phn       
&  DA           &  QSO+star                     &  \nodata\phn  \\
\hline
SDSS~J131339.98+515128.3   &  1.875        &  17.72  &         &                    
&               &                               &               \\ 
SDSS~J131340.02+515129.4   &  1.875        &  17.96  &   1.05  &  Te($VRI$)         
&  LR           &  SDSS lens                    &  4            \\
\hline
SDSS~J131403.48+415203.9   &  1.845        &  18.45  &         &                    
&               &                               &               \\ 
SDSS~J131403.33+415203.7   &  \nodata\phn  &  19.37  &   1.80  &  Te($I$)           
&  \nodata\phn  &  no lens object            &  \nodata\phn  \\
\hline
SDSS~J132405.28+282333.5   &  0.904        &  18.48  &         &                    
&               &                               &               \\ 
SDSS~J132405.19+282331.9   &  \nodata\phn  &  19.69  &   2.03  &  Te($I$)           
&  \nodata\phn  &  no lens object            &  \nodata\phn  \\
\hline
SDSS~J133615.14+285911.8   &  1.423        &  18.89  &         &                    
&               &                               &               \\ 
SDSS~J133614.88+285859.5   &  \nodata\phn  &  19.03  &  12.83  &  WF($i$)           
&  \nodata\phn  &  no lens object            &  \nodata\phn  \\
\hline
SDSS~J134150.74+283207.9   &  1.376        &  18.71  &         &                    
       &               &                        &               \\ 
SDSS~J134150.49+283148.1   &  \nodata\phn  &  19.99  &  20.07  & 
SC($I$)\tablenotemark{e}  &  \nodata\phn  &  no lens object     &  \nodata\phn  \\
\hline
SDSS~J134533.26+113045.2   &  1.355        &  18.75  &         &                    
&               &                               &               \\ 
SDSS~J134532.93+113036.8   &  \nodata\phn  &  19.72  &   9.77  &  \nodata\phn       
&  DA           &  QSO+star                     &  \nodata\phn  \\
\hline
SDSS~J134929.84+122706.9   &  1.722        &  17.46  &         &                    
&               &                               &               \\ 
SDSS~J134930.01+122708.8   &  1.722        &  18.68  &   2.99  &  FO($I$),QU($H$)   
&  DA,FO        &  SDSS lens                    &  1,5  \\
\hline
SDSS~J140530.91+350319.5   &  1.599        &  18.36  &         &                    
&               &                               &               \\ 
SDSS~J140529.53+350328.0   &  0.584        &  18.34  &  19.02  &  \nodata\phn       
&  DA           &  QSO pair                     &  \nodata\phn  \\
\hline
SDSS~J140951.68+384406.1   &  1.675        &  18.51  &         &                    
&               &                               &               \\ 
SDSS~J140950.88+384417.9   &  \nodata\phn  &  18.82  &  15.12  &  \nodata\phn       
&  DA           &  QSO+star                     &  \nodata\phn  \\
\hline
SDSS~J141348.55+475113.4   &  2.175        &  19.00  &         &                    
&               &                               &               \\ 
SDSS~J141348.27+475112.8   &  \nodata\phn  &  18.55  &   2.98  &  NF($H$)           
&  \nodata\phn  &  no lens object            &  \nodata\phn  \\
\hline
SDSS~J141938.37+125227.0   &  1.929        &  19.04  &         &                    
&               &                               &               \\ 
SDSS~J141937.83+125233.7   &  \nodata\phn  &  20.15  &  10.39  &  WF($i$)           
&  \nodata\phn  &  no lens object            &  \nodata\phn  \\
\hline
SDSS~J141951.10+474350.6   &  1.563        &  18.56  &         &                    
&               &                               &               \\ 
SDSS~J141951.14+474357.4   &  \nodata\phn  &  19.28  &   6.76  &  WF($i$)           
&  \nodata\phn  &  no lens object            &  \nodata\phn  \\
\hline
SDSS~J142815.63+095443.5   &  1.467        &  18.48  &         &                    
&               &                               &               \\ 
SDSS~J142815.71+095444.8   &  \nodata\phn  &  19.59  &   1.63  &  Te($I$),NF($H$)   
&  \nodata\phn  &  no lens object            &  \nodata\phn  \\
\hline
SDSS~J143307.88+342315.9   &  1.950        &  19.05  &         &                    
&               &                               &               \\ 
SDSS~J143308.03+342317.1   &  \nodata\phn  &  19.85  &   2.10  &  FO($I$),WF($i$)   
&  \nodata\phn  &  no lens object            &  \nodata\phn  \\
\hline
SDSS~J143624.30+353709.4   &  0.767        &  18.71  &         &                    
&               &                               &               \\ 
SDSS~J143623.21+353707.6   &  \nodata\phn  &  18.78  &  13.41  &  \nodata\phn       
&  DA           &  QSO+star                     &  \nodata\phn  \\
\hline
SDSS~J143715.91+101010.1   &  1.022        &  18.23  &         &                    
&               &                               &               \\ 
SDSS~J143715.87+101008.5   &  \nodata\phn  &  19.12  &   1.76  &  Te($I$)           
&  \nodata\phn  &  no lens object            &  \nodata\phn  \\
\hline
SDSS~J144410.84+304809.6   &  1.731        &  18.17  &         &                    
&               &                               &               \\ 
SDSS~J144410.13+304802.7   &  \nodata\phn  &  18.91  &  11.53  &  WF($i$)           
&  \nodata\phn  &  no lens object            &  \nodata\phn  \\
\hline
SDSS~J151823.05+295925.4   &  1.249        &  18.86  &         &                    
&               &                               &               \\ 
SDSS~J151823.34+295939.8   &  \nodata\phn  &  18.72  &  14.76  &  \nodata\phn       
&  DA           &  QSO+star                       &  \nodata\phn  \\
SDSS~J151823.43+295927.6   &  1.249        &  19.88  &   5.28  &  WF($i$)           
&  DA           &  no lens object, binary QSO  &  \nodata\phn  \\
\hline
SDSS~J152626.52+413135.5   &  2.096        &  19.02  &         &                    
&               &                               &               \\ 
SDSS~J152626.60+413147.6   &  1.100        &  19.44  &  12.06  &  \nodata\phn       
&  DA           &  QSO pair                     &  \nodata\phn  \\
\hline
SDSS~J153937.74+302023.6   &  1.644        &  18.67  &         &                    
&               &                               &               \\ 
SDSS~J153937.10+302017.0   &  1.650        &  19.73  &  10.63  &  \nodata\phn       
&  DA           &  QSO pair                     &  \nodata\phn  \\ 
\hline
SDSS~J154515.93+051713.0   &  2.134        &  18.95  &         &                    
&               &                               &               \\ 
SDSS~J154515.57+051729.0   &  \nodata\phn  &  18.75  &  16.87  &  \nodata\phn       
&  DA           &  QSO+star                     &  \nodata\phn  \\
\hline
SDSS~J155130.62+375421.3   &  1.448        &  18.93  &         &                    
&               &                               &               \\ 
SDSS~J155132.07+375410.9   &  \nodata\phn  &  20.09  &  19.97  &  \nodata\phn       
&  DO           &  QSO+star                     &  \nodata\phn  \\
\hline
SDSS~J155218.09+045635.2   &  1.567        &  18.20  &         &                    
&               &                                 &               \\ 
SDSS~J155217.94+045646.8   &  1.567        &  18.62  &  11.69  &  Te($I$)           
&  DA           &  no lens object, binary QSO  &  \nodata\phn  \\
\hline
SDSS~J160127.53+091255.9   &  1.765        &  18.90  &         &                    
&               &                               &               \\ 
SDSS~J160127.56+091258.9   &  \nodata\phn  &  19.60  &   2.97  &  CI($I$),NF($H$)   
&  \nodata\phn  &  no lens object            &  \nodata\phn  \\
\hline
SDSS~J160602.81+290048.7   &  0.770        &  18.31  &         &                    
&               &                                 &               \\ 
SDSS~J160603.02+290050.9   &  0.770        &  18.38  &   3.45  &  SP($i$)           
&  DA           &  no lens object, binary QSO  &  6  \\
\hline
SDSS~J160614.69+230518.0   &  1.206        &  18.89  &         &                    
      &         &                               &               \\ 
SDSS~J160614.80+230518.1   &  \nodata\phn  &  19.27  &   1.37  & 
Te($I$),NF($H$),UF($K$)  &  WF     &  different SED, not QSO       &  \nodata\phn 
\\
\hline
SDSS~J161055.42+354404.7   &  1.549        &  18.96  &         &                    
&               &                               &               \\ 
SDSS~J161056.18+354418.3   &  \nodata\phn  &  19.40  &  16.36  &  CI($I$),WF($i$)   
&  \nodata\phn  &  no lens object            &  \nodata\phn  \\
\hline
SDSS~J161526.64+264813.7   &  2.179        &  18.40  &         &                    
&               &                               &               \\ 
SDSS~J161526.35+264819.2   &  \nodata\phn  &  19.56  &   6.74  &  \nodata\phn       
&  DA           &  QSO+star                     &  \nodata\phn  \\ 
SDSS~J161527.21+264813.7   &  \nodata\phn  &  19.63  &   7.54  &  \nodata\phn       
&  DA           &  QSO+star                     &  \nodata\phn  \\ 
\hline
SDSS~J162919.93+231919.9   &  0.852        &  18.50  &         &                    
&               &                               &               \\ 
SDSS~J162919.09+231933.4   &  \nodata\phn  &  18.68  &  17.81  &  \nodata\phn       
&  DA           &  QSO+star                     &  \nodata\phn  \\ 
\hline
SDSS~J163520.04+205225.1   &  1.775        &  19.03  &         &                    
&               &                                                &              \\ 
SDSS~J163519.51+205213.9   &  1.775        &  20.07  &  13.61  &  Te($I$),CI($I$)   
&  DA,DO        &  no lens object, different SED, binary QSO  &  \nodata\phn  \\
\hline
SDSS~J164212.07+220049.0   &  1.552        &  18.55  &         &                    
&               &                               &               \\ 
SDSS~J164211.45+220038.0   &  \nodata\phn  &  19.37  &  14.08  &  FO($I$)           
&  \nodata\phn  &  no lens object            &  \nodata\phn  \\ 
\hline
SDSS~J164302.96+132738.2   &  1.526        &  18.88  &         &                    
&               &                               &               \\ 
SDSS~J164302.69+132724.5   &  \nodata\phn  &  19.35  &  14.28  &  CI($I$),WF($i$)   
&  \nodata\phn  &  no lens object            &  \nodata\phn  \\
\hline
SDSS~J164655.14+194300.8   &  1.951        &  19.02  &         &                    
&               &                               &               \\ 
SDSS~J164656.25+194310.5   &  \nodata\phn  &  19.61  &  18.34  &  WF($i$)           
&  \nodata\phn  &  no lens object            &  \nodata\phn  \\
\hline
SDSS~J164723.58+203314.4   &  0.814        &  18.24  &         &                    
&               &                               &               \\ 
SDSS~J164723.80+203308.6   &  \nodata\phn  &  18.94  &   6.55  &  Te($I$)           
&  \nodata\phn  &  no lens object            &  \nodata\phn  \\
\hline
SDSS~J165156.72+280036.9   &  0.862        &  18.65  &         &                    
&               &                               &               \\ 
SDSS~J165156.98+280036.4   &  \nodata\phn  &  19.72  &   3.38  &  Te($I$)           
&  \nodata\phn  &  QSO+galaxy                   &  \nodata\phn  \\
\hline
SDSS~J165326.37+193326.5   &  1.534        &  19.10  &         &                    
&               &                               &               \\ 
SDSS~J165326.23+193316.3   &  \nodata\phn  &  18.27  &  10.45  &  WF($i$)           
&  \nodata\phn  &  no lens object            &  \nodata\phn  \\
\hline
SDSS~J165502.02+260516.5   &  1.892        &  17.64  &         &                    
&               &                                     &               \\ 
SDSS~J165501.32+260517.5   &  1.881        &  17.77  &   9.55  &  \nodata\phn       
&  \nodata\phn  &  SDSS QSO, different SED, QSO pair  &  \nodata\phn  \\ 
\hline
SDSS~J165609.48+250857.5   &  2.165        &  18.98  &         &                    
&               &                               &               \\ 
SDSS~J165608.98+250853.8   &  \nodata\phn  &  19.03  &   7.84  &  CI($I$)           
&  DO           &  QSO+star                     &  \nodata\phn  \\ 
\hline
SDSS~J170438.29+212149.3   &  1.438        &  18.80  &         &                    
&               &                               &               \\ 
SDSS~J170438.69+212202.1   &  \nodata\phn  &  19.36  &  13.95  &  CI($I$),WF($i$)   
&  \nodata\phn  &  no lens object            &  \nodata\phn  \\
\hline  
SDSS~J170555.21+214801.8   &  1.249        &  18.93  &         &                    
&               &                               &               \\ 
SDSS~J170555.06+214801.7   &  \nodata\phn  &  19.84  &   2.18  &  UF($K$)           
&  \nodata\phn  &  no lens object            &  \nodata\phn  \\ 
\hline 
SDSS~J221953.73$-$010037.5 &  1.700        &  18.97  &         &                    
&               &                               &               \\ 
SDSS~J221953.54$-$010032.1 &  \nodata\phn  &  19.56  &   6.14  &  UF($K$)           
&  \nodata\phn  &  no lens object            &  \nodata\phn  \\
\hline
SDSS~J234348.33+000202.8   &  1.812        &  19.08  &         &                    
&               &                               &               \\ 
SDSS~J234348.49+000203.4   &  \nodata\phn  &  20.14  &   2.38  &  UF($K$)           
&  \nodata\phn  &  no lens object            &  \nodata\phn  \\
\enddata
\tablecomments{Two candidates, SDSS~J080623.706+200631.89 and
SDSS~J145307.064+331950.54, 
that are identified by the morphological selection algorithm as well, are listed in 
Table \ref{tab:can_mor}.}
\tablenotetext{a}{Redshifts from the SDSS DR5 quasar catalog.} 
\tablenotetext{b}{$i$-band PSF magnitudes with Galactic extinction
   corrections from the SDSS DR5 quasar catalog.}  
\tablenotetext{c}{Image separations ($\theta_{\rm SDSS}$) in units of
   arcsec between the two components from the SDSS imaging data.} 
\tablenotetext{d}{Instruments (and filters) used for the follow-up observations. 
   DA: DIS III at ARC 3.5m, NF: NICFPS at ARC 3.5m, SP: SPIcam at ARC 3.5m,
   UF: UFTI at UKIRT, 
   Te: Tek2k CCD at UH88, 8k: UH8k at UH88, WF: WFGS2 at UH88, QU: QUIRC at UH88,
   FO: FOCAS at Subaru, 
   LR: LRIS at Keck, 
   DO: DOLORES at TNG 3.6m, 
   CI: CCD Imager at KPNO 2.1m.
   }
\tablenotetext{e}{The data are obtained from the SMOKA \citep{baba02}.} 
\tablerefs{
   (1) \citealt{hennawi06a};
   (2) \citealt{oguri05}; 
   (3) \citealt{oguri08b};
   (4) \citealt{ofek07};
   (5) \citealt{kayo09};
   (6) \citealt{myers08}.
   }
\end{deluxetable}

\clearpage

\begin{deluxetable}{lccccrccll}
\tabletypesize{\footnotesize}
\rotate
\tablecaption{DR5 Statistical Sample \label{tab:lens_dr5stat}}
\tablewidth{0pt}
\tablehead{ 
\colhead{Object} & 
\colhead{$N_{\rm img}$} & 
\colhead{$z_s$\tablenotemark{a}} &
\colhead{$z_l$\tablenotemark{b}} &
\colhead{$M_I$\tablenotemark{c}} &
\colhead{$\theta_{\rm max}$\tablenotemark{d}} & 
\colhead{$f_{I}$ \tablenotemark{e}} & 
\colhead{Source} & 
\colhead{Comment} & 
\colhead{Ref.} 
}
\startdata
SDSS~J0246$-$0825  &  2  &  1.682  &  0.723        &  $-$21.9      &  1.04  &  0.34
&  DR3  &  SDSS lens                &  1,2,3 \\ 
SDSS~J0746+4403    &  2  &  2.003  &  0.513        &  $-$22.6      &  1.08  &  0.97
&  DR5  &  SDSS lens                &  4,5  \\ 
SDSS~J0806+2006    &  2  &  1.540  &  0.573        &  $-$22.3      &  1.40  &  0.76
&  DR5  &  SDSS lens                &  3,6  \\ 
SDSS~J0913+5259    &  2  &  1.377  &  0.830        &  $-$24.5      &  1.14  &  0.70
&  DR3  &  known lens SBS~0909+523  &  2,7,8 \\
SDSS~J0924+0219    &  4  &  1.524  &  0.394        &  $-$23.0      &  1.78  &  0.43
&  DR3  &  SDSS lens                &  2,9,10,11 \\ 
SDSS~J1001+5027    &  2  &  1.838  &  \nodata\phn  &  \nodata\phn  &  2.86  &  0.72
&  DR3  &  SDSS lens                &  2,12  \\ 
SDSS~J1001+5553    &  2  &  1.405  &  0.390        &  $-$24.4      &  6.17  &  0.94
&  DR3  &  known lens Q0957+561     &  2,13,14 \\
SDSS~J1004+4112    &  5  &  1.732  &  0.680        &  $-$23.9      & 14.62  &  0.23
&  DR3  &  SDSS lens                &  2,15  \\ 
SDSS~J1021+4913    &  2  &  1.720  &  \nodata\phn  &  \nodata\phn  &  1.14  &  0.40
&  DR3  &  SDSS lens                &  2,16 \\ 
SDSS~J1055+4628    &  2  &  1.249  &  \nodata\phn  &  \nodata\phn  &  1.19  &  0.33
&  DR5  &  SDSS lens                &  5 \\ 
SDSS~J1118+0745    &  4  &  1.720  &  0.311        &  $-$22.2      &  2.43  &  0.25
&  DR5  &  known lens PG1115+080    &  17,18,19\\ 
SDSS~J1206+4332    &  2  &  1.789  &  \nodata\phn  &  \nodata\phn  &  2.90  &  0.74
&  DR5  &  SDSS lens                &  12   \\ 
SDSS~J1216+3529    &  2  &  2.012  &  \nodata\phn  &  \nodata\phn  &  1.49  &  0.41
&  DR5  &  SDSS lens                &  20   \\ 
SDSS~J1226$-$0006  &  2  &  1.121  &  0.517        &  $-$22.5      &  1.24  &  0.45
&  DR3  &  SDSS lens                &  2,21  \\ 
SDSS~J1313+5151    &  2  &  1.875  &  0.194        &  $-$21.6      &  1.24  &  0.92
&  DR5  &  SDSS lens                &  22   \\ 
SDSS~J1332+0347    &  2  &  1.445  &  0.191        &  $-$21.6      &  1.14  &  0.70
&  DR3  &  SDSS lens                &  2,24 \\
SDSS~J1335+0118    &  2  &  1.570  &  0.440        &  $-$22.2      &  1.56  &  0.37
&  DR3  &  SDSS lens                &  2,21,24 \\ 
SDSS~J1353+1138    &  2  &  1.629  &  \nodata\phn  &  \nodata\phn  &  1.41  &  0.40
&  DR5  &  SDSS lens                &  6    \\ 
SDSS~J1524+4409    &  2  &  1.210  &  0.320        &  $-$22.7      &  1.67  &  0.56
&  DR3  &  SDSS lens                &  2,20 \\ 
\enddata
\tablenotetext{a}{Source redshifts from follow-up observations} 
\tablenotetext{b}{Measured lens redshifts.} 
\tablenotetext{c}{Absolute magnitudes of the detected (brightest) lens galaxies. 
                  The combinations of evolution- and
                  $K$-corrections are included \citep{poggianti97}.} 
\tablenotetext{d}{Maximum image separations in units of arcsec.}
\tablenotetext{e}{Flux ratios between the brightest lensed image and the 
                  farthest lensed image from the brightest image, in
		  the $I$-band images.} 
\tablerefs{(1) \citealt{inada05}; 
           (2) Paper II;
           (3) \citealt{eigenbrod07};
           (4) \citealt{inada07};
           (5) \citealt{kayo09};
           (6) \citealt{inada06};
           (7) \citealt{oscoz97};
           (8) \citealt{lubin00};
           (9) \citealt{inada03a};
           (10) \citealt{ofek07};
           (11) \citealt{eigenbrod06a};
           (12) \citealt{oguri05};
           (13) \citealt{walsh79};
           (14) \citealt{young80};
           (15) \citealt{inada03b};          
           (16) \citealt{pindor06};
           (17) \citealt{weymann80};
           (18) CASTLES webpage (C.~S.~Kochanek et al.,
http://cfa-www.harvard.edu/castles/.);
           (19) \citealt{kundic97};
           (20) \citealt{oguri08b};
           (21) \citealt{eigenbrod06b};
           (22) \citealt{ofek07};
           (23) \citealt{morokuma07};
           (24) \citealt{oguri04}.
           }
\end{deluxetable}

\clearpage

\begin{deluxetable}{lcccrcclll}
\tabletypesize{\footnotesize}
\rotate
\tablecaption{DR5 Additional Lensed Quasars\label{tab:lens_dr5add}}
\tablewidth{0pt}
\tablehead{ 
\colhead{Object} & 
\colhead{$N_{\rm img}$} & 
\colhead{$z_s$\tablenotemark{a}} &
\colhead{$z_l$\tablenotemark{b}} &
\colhead{$\theta$\tablenotemark{c}} & 
\colhead{$f_{I}$ \tablenotemark{d}} & 
\colhead{Source} & 
\colhead{Comment} & 
\colhead{Reject\tablenotemark{e}} & 
\colhead{Ref.} 
}
\startdata
SDSS~J0134$-$0931  &  5  &  2.216  &  0.765        &  0.68  &  0.03  &  DR3  & known
lens PMN~J0134$-$0931  &  $z>2.2$, $i>19.1$, $\theta<1''$         &  1,2,3,4 \\  
SDSS~J0145$-$0945  &  2  &  2.719  &  0.491        &  2.23  &  0.15  &  DR3  & known
lens Q0142$-$100       &  $z>2.2$, $|{\Delta} I|>1.25$            &  4,5,6 \\ 
SDSS~J0820+0812    &  2  &  2.024  &  0.803        &  2.30  &  0.17  &  DR5  & known
lens ULAS~J0820+0812   &  $|{\Delta} I|>1.25$                     &  7  \\ 
SDSS~J0813+2545    &  4  &  1.500  &  \nodata\phn  &  0.91  &  0.06  &  DR3  & known
lens HS~0810+2554      &  $\theta<1''$                            &  4,8 \\
SDSS~J0819+5356    &  2  &  2.237  &  0.294        &  4.04  &  0.23  &  DR5  & SDSS
lens                    &  $z>2.2$, $|{\Delta} I|>1.25$            &  9  \\
SDSS~J0832+0404    &  2  &  1.115  &  0.659        &  1.98  &  0.22  &  DR3  & SDSS
lens                    &  $|{\Delta} I|>1.25$                     &  4,10 \\
SDSS~J0903+5028    &  2  &  3.584  &  0.388        &  2.80  &  0.46  &  DR3  & SDSS
lens                    &  $z>2.2$, $i>19.1$                       &  4,11 \\
SDSS~J0911+0550    &  4  &  2.800  &  0.769        &  3.26  &  0.41  &  DR3  & known
lens RX~J0911+0551     &  $z>2.2$                                 &  4,12,13 \\
SDSS~J1004+1229    &  2  &  2.650  &  0.950        &  1.54  &  0.09  &  DR5  & known
lens J1004+1229        &  $z>2.2$, $i>19.1$, $|{\Delta} I|>1.25$  &  14,15 \\ 
SDSS~J1138+0314    &  4  &  2.442  &  0.445        &  1.44  &  0.35  &  DR3  & SDSS
lens                    &  $z>2.2$                                 &  4,16 \\
SDSS~J1155+6346    &  2  &  2.890  &  0.176        &  1.83  &  0.50  &  DR3  & SDSS
lens                    &  $z>2.2$                                 &  4,17 \\  
SDSS~J1322+1052    &  2  &  1.716  &  \nodata\phn  &  2.00  &  0.21  &  DR5  & SDSS
lens                    &  $|{\Delta} I|>1.25$                     &  10  \\ 
SDSS~J1349+1227    &  2  &  1.722  &  \nodata\phn  &  3.01  &  0.30  &  DR5  & SDSS
lens                    &  $|{\Delta} I|>1.25$                     &  18  \\ 
SDSS~J1406+6126    &  2  &  2.126  &  0.271        &  1.98  &  0.58  &  DR3  & SDSS
lens                    &  $i>19.1$                                &  4,19 \\
SDSS~J1633+3134    &  2  &  1.511  &  \nodata\phn  &  0.66  &  0.30  &  DR3  & known
lens FBQ~1633+3134     &  $\theta<1''$, $|{\Delta} I|>1.25$       &  4,20 \\
SDSS~J1650+4251    &  2  &  1.547  &  \nodata\phn  &  1.20  &  0.17  &  DR3  & SDSS
lens                    &  $|{\Delta} I|>1.25$                     &  4,21 \\
SDSS~J2343$-$0050  &  2  &  0.788  &  \nodata\phn  &  1.51  &  0.85  &  DR5  & known
lens ULAS~J2343$-$0050 &  $i>19.1$                                &  22  \\ 
\enddata
\tablecomments{See text for the selection of each lensed quasar.} 
\tablenotetext{a}{Source redshifts from follow-up observations} 
\tablenotetext{b}{Measured lens redshifts.} 
\tablenotetext{c}{Maximum image separations in units of arcsec.}
\tablenotetext{d}{Flux ratios between the brightest lensed image and the 
                  farthest lensed image from the brightest image, in
                  the $I$-band images.} 
\tablenotetext{e}{The reason that each lens is excluded from the
  statistical sample.}  
\tablerefs{(1) \citealt{winn02};
           (2) \citealt{gregg02};
           (3) \citealt{hall02};
           (4) Paper II;
           (5) \citealt{surdej87};
           (6) \citealt{eigenbrod07};
           (7) \citealt{jackson09};
           (8) \citealt{reimers02};
           (9) \citealt{inada09}; 
           (10) \citealt{oguri08b};
           (11) \citealt{johnston03};
           (12) \citealt{bade97};
           (13) \citealt{kneib00};
           (14) \citealt{lacy02};
           (15) CASTLES webpage (C.~S.~Kochanek et al.,
http://cfa-www.harvard.edu/castles/.).
           (16) \citealt{eigenbrod06b};
           (17) \citealt{pindor04};
           (18) \citealt{kayo09};
           (19) \citealt{inada07};
           (20) \citealt{morgan01};
           (21) \citealt{morgan03};
           (22) \citealt{jackson08}.
           }
\end{deluxetable}

\clearpage

\clearpage

\end{document}